\ifx\pdfoutput\undefined\else\ifx\XeTeXrevision\undefined\pdfoutput=1\fi\fi 
\RequirePackage[T1]{fontenc}
\documentclass[conference]{IEEEtran}
\usepackage{amsmath,amssymb}
\usepackage{graphicx}
\usepackage{booktabs,tabularx}
\usepackage{listings}

\usepackage{xcolor}
\usepackage{cite}
\usepackage[hyphens]{url}
\definecolor{navy}{HTML}{12264A}
\definecolor{blue}{HTML}{1660D2}
\definecolor{teal}{HTML}{118F83}
\definecolor{pale}{HTML}{F3F2FF}
\usepackage[colorlinks=true,allcolors=blue]{hyperref}
\hypersetup{pdftitle={Can Coding Agents Migrate to Post-Quantum Cryptography?},pdfauthor={Abdulmalik Alquwayfili}}
\ExplSyntaxOn
\tl_replace_all:Nnn \abstract {---} {:~}
\tl_replace_all:Nnn \IEEEkeywords {---} {:~}
\ExplSyntaxOff
\graphicspath{{figures/}}

\title{Can Coding Agents Migrate to\\Post-Quantum Cryptography?}
\author{\IEEEauthorblockN{Abdulmalik Alquwayfili}
\IEEEauthorblockA{\textit{Saudi Data \& AI Authority (SDAIA), Saudi Arabia}\\
\href{mailto:aalquwayfili@ncai.gov.sa}{\textcolor{black}{aalquwayfili@ncai.gov.sa}}}}
\begin{document}
\bstctlcite{bibliographycontrol}
\maketitle

\begin{abstract}
A program migrated to post-quantum cryptography can verify its own
signatures while producing keys or signatures that another implementation
rejects. We introduce a contract-based task for migrating a
Go file signer from RSA to ML-DSA-44, and compare coding agents with and
without structured checker feedback. Both conditions receive the contract,
compiler, documentation, and OpenSSL. Across 160 attempts in four
local-agent configurations, twelve final patches pass local verification
but fail external requirements. Checker access does not increase the
observed completion rate in any comparison. Recorded traces show
unresolved defects and checks invoked only after a patch is correct.
Serving settings also affect completion: reducing only Qwen3.8's context
window from 128K to 32K lowers full passes from 36/40 to 4/40.
Four exploratory trials using GPT-6 Astra through Codex and Claude
Fable 5.1 through Claude Code pass all 40 checks, including both baselines. These findings support evaluating
interoperability separately from local agreement and reporting harness
and serving limits alongside agent results.

\end{abstract}


\section{Introduction}
Migrating a signature application to ML-DSA requires more than replacing
the signing function.\footnote{Code: \url{https://github.com/aalquwayfili/pqc-migration}.}
The application must read and write usable keys,
sign the intended message, and distinguish an invalid signature from an
input it cannot process. FIPS~204 specifies ML-DSA~\cite{fips204}; the
application must also define the interface and data formats that other
software will use.

A local sign--verify test can miss a mistake shared by both operations.
In two recorded attempts, the application wraps packed key bytes in PEM
and reads them back with a matching loader. Local signing and verification
pass, but OpenSSL rejects the files because the required key structures
are missing.

\begin{figure}[t]
  \centering
  \includegraphics[width=\columnwidth,trim=0 9bp 0 0,clip]{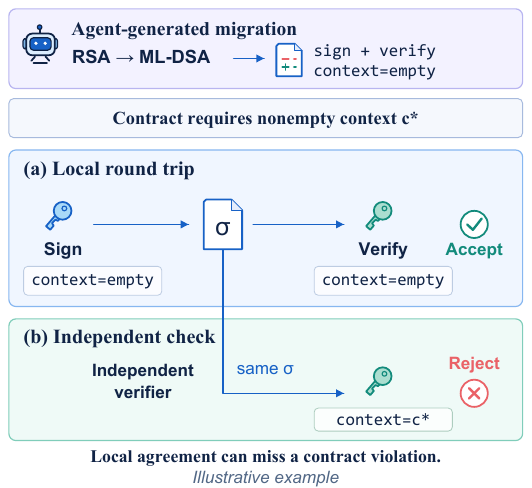}
  \caption{Local signing and verification agree on the wrong context; an independent verifier using the required context rejects the signature.}
  \label{fig:teaser}
\end{figure}

Figure~\ref{fig:teaser} illustrates the same problem with context:
signing and verification agree on an empty context, but an independent
verifier rejects the signature under the required nonempty context.
ML-DSA permits empty contexts~\cite{fips204,opensslmldsa}; the error is
ignoring this application's requirement.

When a coding agent writes both a migration and its tests, a shared
mistake can go undetected. Compilation catches type and API errors;
verification by a separate implementation tests agreement on message
bytes, context, and key encoding. The question is whether diagnostics
from these checks help the agent repair its mistakes.

We compare two ways of giving an agent feedback during migration.
Both conditions receive the same application, contract, documentation,
compiler, and OpenSSL installation. The baseline can write tests and check
interoperability itself. The feedback condition also has optional access to structured diagnostics from an external checker. Both have the same editing budget, and a common
final evaluator checks their patches on fresh inputs. This comparison asks
whether structured diagnostics help beyond access to the verification tool.

The task uses CIRCL and specifies persisted keys, a fixed signing context,
and separate responses for invalid signatures and malformed inputs.
We evaluate four local-agent configurations with 20 paired
baseline--feedback attempts each, plus four exploratory trials with
Codex and Claude Code. Twelve patches pass the local round trip while
failing external requirements; no comparison shows a higher completion
rate with checker access.

The study contributes a reproducible migration task and checker,
four comparisons of optional feedback, an ablation that changes only
Qwen3.8's served context window, and recorded examples of incomplete
repairs. Within each feedback comparison, the model and environment
remain fixed.

\section{Related Work}
\label{sec:related}
\textbf{Post-quantum migration.}
Strauss et al. examine cryptographic use in Android applications and
explore LLM-assisted migration, reporting difficulties with dependencies
and changes across files~\cite{strauss2025readiness}.
Pallar\'{e}s de Bonrostro et al. evaluate zero-shot and fine-tuned models
on paired Python fragments and examine migrations in open-source
projects~\cite{pallares2026migration}. Their functional tests give each migration category executable
acceptance criteria. The illustrated signature
validator checks a local sign--verify round trip before and after
migration. This establishes local agreement in each branch; it does not
by itself test whether the migrated application exchanges keys and
signatures with another implementation.

Zhang's AQuA vision places verification alongside cryptographic discovery
and refactoring~\cite{zhang2026aqua}. Its proposed workflows combine
regression tests, differential and metamorphic testing, and selected
analyses of side-channel behavior. AQuA sets out a research agenda for
migration tools. Our study evaluates a concrete file-signer migration
that preserves the interface and specifies messages, context, persisted
keys, and rejections.

\textbf{Independent checks.}
A test oracle judges whether observed behavior is correct. Barr et al.
survey oracles derived from specifications, contracts, and other
evidence~\cite{barr2015oracle}. Differential testing uses disagreement
between implementations to expose possible faults, as Csmith does for
compiler outputs~\cite{yang2011csmith}. In cryptographic software,
Frankencerts generates certificates from mutated components and compares
validation decisions across SSL/TLS implementations~\cite{brubaker2014frankencerts}.
MLCerts learns certificate representations with generative language
models and samples synthetic inputs for differential
testing~\cite{paracha2026mlcerts}.
Our checker exchanges keys and signatures across implementations under a
declared application contract. Because ML-DSA supports randomized signing, the relevant
observations are successful key loading, acceptance of valid signatures,
and rejection of invalid ones, rather than identical signature bytes.

\textbf{Coding agents.}
SWE-bench evaluates repository patches using tests for real GitHub
issues~\cite{jimenez2024swebench}. SWE-agent lets models navigate
repositories, edit files, and run programs~\cite{yang2024sweagent}.
Suresh et al. give a post-task memory curator read-only environment tools
to check lessons before reuse~\cite{suresh2026grounding}.
Kiecker et al. study iterative repair across code generation, test
generation, and translation, finding that most gains occur in the first
few repair iterations~\cite{kiecker2026repairloops}. Their workflows return
validation failures to the model. Our comparison leaves checker
invocation to the agent, distinguishing access to diagnostics from a
workflow that delivers them after each failed validation.

\textbf{Security evaluation.}
BaxBench uses functional tests and executable exploits to show that
generated backends can pass functional tests while remaining
vulnerable~\cite{vero2025baxbench}. MPC-Patch-Bench examines repairs in
multi-party computation software~\cite{zhang2026mpcpatch}. Its verifier
combines comparisons against plaintext oracles with static checks for
security violations. In the reported protocol, patches that pass the
functional tests undergo this additional evaluation.

Chai uses agent-guided differential testing to discover cryptographic
misuse and traces library discrepancies into downstream
applications~\cite{villa2026chai}. It replays and minimizes discrepancies,
classifies their consequences, and retains human review before reporting.
Its use of independent implementations to guide an agent is closely
related to our feedback setting.

We examine whether access to these diagnostics during repair
helps an agent meet a migration contract within a fixed budget. The
comparison holds the agent, editing interface, and development tools
fixed. Both conditions can write tests and use OpenSSL; one also receives
checker feedback, and a separate evaluator assesses their final patches.
Our outcome is contract compliance: a context or key-format error need
not be exploitable.

\section{Method}
\label{sec:method}
\begin{figure*}[t]
  \centering
    \includegraphics[width=\textwidth,trim=0 38bp 0 0,clip]{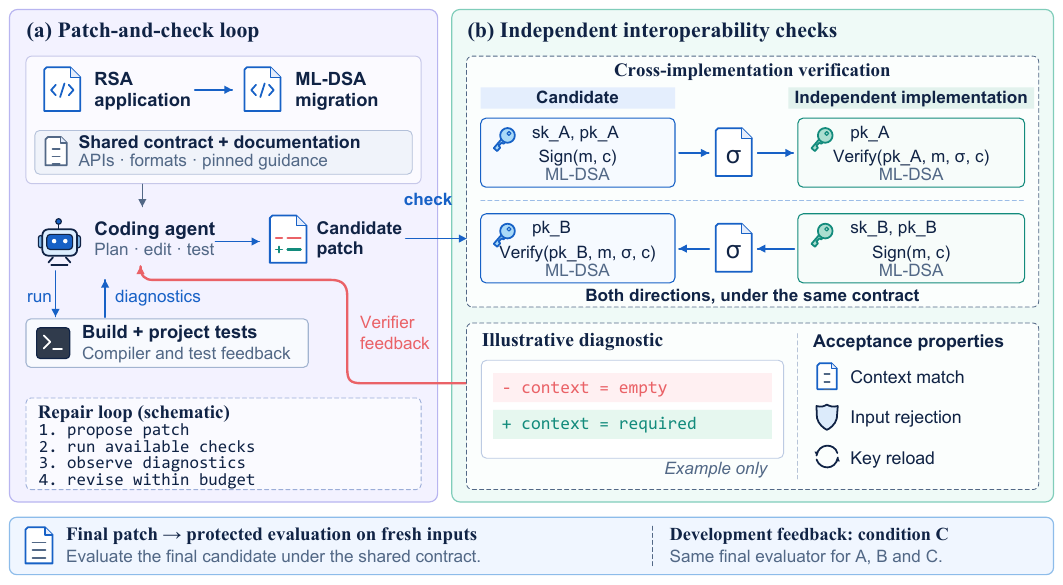}
  \par\vspace{3pt}
  \begingroup
  \setlength{\fboxsep}{5pt}
  \colorbox{blue!6}{\parbox{\dimexpr\textwidth-2\fboxsep\relax}{%
    \centering\footnotesize Final patches from both conditions receive
    the same evaluation on fresh inputs.}}
  \endgroup
  \caption{Patch-and-check loop with independent interoperability feedback.
  Development tests are agent-written; the context diagnostic is illustrative.}
  \label{fig:method}
\end{figure*}

The agent edits the application to meet a written contract. An external checker tests the patch through its public interface (Figure~\ref{fig:method}). During development, its
diagnostics are available only in the feedback condition.

\subsection{Task}
The reference is a file signer with three entry points:
\texttt{keygen}, \texttt{sign}, and \texttt{verify}. Keys persist on disk and signatures
are stored separately from the input file. The migration preserves
these entry points while changing the signing primitive. The legacy
version computes a SHA-256 digest and signs it with RSA-2048 using
PKCS\#1~v1.5. The migrated version passes the original file bytes to
Pure ML-DSA-44, without an application-level prehash.

Both signing and verification use \texttt{filesigner/v1} as their context,
a requirement introduced by the migration. The reference passes this value
directly to CIRCL's signing and verification functions~\cite{circlmldsa}.

Persisted keys must be usable by the external implementation without
conversion. Public keys use PEM-wrapped SubjectPublicKeyInfo; private
keys use PEM-wrapped PKCS\#8. The reference writes the ML-DSA seed and
expanded key together and checks their agreement when loading that
form. Signing uses the library's randomized mode, so acceptance depends
on verification, not a fixed signature byte string.

\subsection{Interoperability}
The checker invokes application binaries as subprocesses and exchanges
files with OpenSSL. It imports no application code. Let $A$ denote the
candidate and $B$ the independent implementation, with signing and
verification operations $S$ and $V$. For message $m$ and required
context $c^*$, the two directions require
\begin{align}
 V_B(pk_A,m,S_A(sk_A,m,c^*),c^*) &= 1,\\
 V_A(pk_B,m,S_B(sk_B,m,c^*),c^*) &= 1.
\end{align}
Each side generates its key pair. Loading external keys also tests
the interchange format.
The reference pins CIRCL v1.6.5 and OpenSSL~3.5.8; OpenSSL exposes an
explicit ML-DSA context parameter~\cite{opensslmldsa}.

Rejection tests modify the message,
use a different key, change or remove the context, and corrupt the
signature. The checker also tests truncated signatures and unparseable
keys. Persistence checks write and reload keys before signing and
verifying again. 

\subsection{Outcomes}
The contract distinguishes a cryptographic rejection from malformed
input and execution failure. The application returns exit code~1 for
an invalid signature and code~3 for malformed input. The checker compares
the observed response with the required response for each case; an
expected rejection counts as a passed check. It records contract failures,
missing dependencies, and execution errors separately.

The checker identifies timeouts, signals, and Go runtime failures before
interpreting an exit code. This prevents a crash from counting as a correct
rejection. Validation controls exercise context mistakes, unconditional
acceptance, incorrect error codes, and faults in the checker. Correct
implementations are included as controls for false failures. The v4 reference passes all 40 checks, and the evaluator detects all
16 validation controls, including through the isolated scoring path.
These validate the study's machinery and are excluded from agent results.

\subsection{Feedback}
In the feedback condition, the agent issues \texttt{check} as a standalone
action. The host returns build status and named outcomes, including expected
and observed responses where they differ. The agent can use these diagnostics
to revise its patch within the remaining budget. Final evaluation follows
editing and supplies no repair feedback.

\section{Evaluation}
\label{sec:evaluation}

\subsection{Comparison}
We compare final correctness, errors missed by local checking, and
checking cost. Both conditions receive the full contract, RSA source,
documentation, compiler, and OpenSSL. The baseline can write interoperability
tests; feedback also provides \texttt{check}. Development and final builds
share an offline proxy containing only CIRCL and \texttt{golang.org/x/sys}.
OpenSSL is available during development and to the final evaluator,
but absent from the scored candidate's runtime.

All local configurations use mini-swe-agent 2.4.6~\cite{yang2024sweagent} with LiteLLM 1.100.1.
The first comparison uses Qwen3-Coder-30B-A3B-Instruct~\cite{qwen3coder} through
vLLM 0.24.0, with float16 weights and a
32,768-token context. Sampling uses temperature 0.7, top-$p$ 0.8,
top-$k$ 20, and repetition penalty 1.05. Each local attempt allows 60 model
requests, one hour including model and checking time, and at most 8,192
output tokens per response. The prompt describes a limit of 60 commands,
but the harness counts model requests, including rejected responses.
An attempt can therefore exhaust its budget before executing 60 commands;
we report executed actions separately.

We run Qwen3.8-27B~\cite{qwen38} in Q4\_K\_M precision through
llama.cpp, revision \texttt{4a89937}, with a matching Q8\_0 multi-token
prediction (MTP) draft head from ggml-org~\cite{qwen38gguf} and a
131,072-token context. Thinking and
retention of prior reasoning are enabled. Sampling uses temperature 1.0,
top-$p$ 0.95, top-$k$ 20, and min-$p$ 0; the request timeout is 1,500
seconds. The task, prompt, parser, development tools, final evaluator,
and request, time, and output budgets remain the same. A second Qwen3.8
batch changes only the served context window to 32,768 tokens. We retain
the same weights, draft head, inference engine, sampling, and paired seeds.

The fourth configuration uses Nemotron 3.5 Lightning (30B parameters,
3B active)~\cite{nemotron}
in Q4\_0 precision with a matching Q4\_0 MTP head~\cite{nemotrongguf}.
It uses the same llama.cpp revision, 131,072-token window, agent harness,
tools, and budgets as Qwen3.8. Thinking and retention of prior reasoning
remain enabled. Sampling uses temperature 1.0, top-$p$ 0.95, top-$k$ 0,
and min-$p$ 0.

Each comparison uses 20 paired seeds, one attempt per condition and
seed, in baseline--feedback order. Qwen3-Coder attempts run sequentially
on two RTX 8000 GPUs. The Qwen3.8 and Nemotron batches run one server per GPU and two attempts
at a time. Odd and even seeds use different GPUs; each seed's baseline
and feedback attempts use the same GPU. No unrelated benchmark traffic
shares the servers, although concurrent workers share the host CPU.

The Qwen3.8 128K comparison follows the Qwen3-Coder and native trials.
Its setup pilots exposed timeout and context limits, which informed
the serving configuration. All setup pilots and serving probes are
excluded. The 32K ablation and Nemotron batch follow the 128K comparison.
The shared build/test/repair instruction predates all four
counted comparisons. Settings and identities were frozen before each
batch, using the same analysis rules. Comparisons across models also
change precision, inference engine, context, sampling, and load; they
are not controlled estimates of model capability.

Each attempt starts in a fresh workspace. The agent's final patch is the
workspace left when it submits or reaches a stopping limit. An agent exit
on a model or format error is also scored. Repetitions measure variation on this
task, rather than adding new migration tasks. Each run records its prompt,
model revision, sampling settings, seed, limits, and stopping reason.

\subsection{Native agents}
After the Qwen3-Coder study, we run one baseline and one feedback attempt with
each of two native coding agents: Codex CLI 0.154.0 with
\texttt{gpt-6-astra}, and Claude Code 2.1.272 with
\texttt{claude-fable-5-1}, both at high reasoning effort. They receive the
same task, starting source, offline dependencies, and final evaluator.
Within each agent, prompts differ only in checker access. External
network access is restricted to the model providers.

The limits remain one hour and 60 tool actions. Native file edits and
shell commands count as actions; these units differ from the local
comparisons' executed shell steps. The CLIs also manage context differently.
Claude commands have a 120-second timeout; Codex chooses command timeouts
and uses an auxiliary approval model. Native trial identities are labels,
not controlled sampling seeds. We report these runs as individual
agent-configuration cases, separately from the paired comparisons.

\subsection{Final checks}
Final evaluation rebuilds a snapshot of the selected patch and tests it
with fresh inputs. The source snapshot is mounted read-only; compilation
writes the binary to a separate output directory. Both conditions use the
same evaluator and a 30-second limit per candidate invocation. The scored program receives
the files needed for each operation. Isolation probes check that the agent and scored program cannot access
the reference, checker, or other protected files.

The external checker includes a local sign--verify round trip, allowing
the same final patch to be classified by local and external outcomes.
A patch may pass the local check and fail an interoperability requirement.
The starting application has no tests; agent-written tests are retained
in the development trace and reported separately.

\subsection{Measures}
The primary outcome is the share of valid counted attempts passing every
final check. Failed builds and unsuccessful repairs stay in the denominator.
We report counts, 95\% Wilson intervals, the difference in percentage points,
and paired-seed outcomes: both pass, baseline only, feedback only, or neither.

Pre-action infrastructure failures and independently verified external
outages are logged before final scoring and retried from the starting state
with the same seed. Interruptions and replacements are disclosed; candidate
failures remain scored.

Execution errors, missing dependencies, and cryptographic rejections remain
distinct. A local pass with an external failure requires an observed failed
requirement. Missing checks remain unassessed. We retain raw verdicts but
do not infer rejection defects when required test inputs could not be constructed.

Per-property results cover messages, context, keys, persistence, and
rejections. Repair costs are median and interquartile range of executed
actions, development checker calls, and elapsed time. Checking consumes
both conditions' time budgets.

Prompt growth is measured from each response's reported input tokens,
including cached inputs. Claude events sharing a response identifier
count as one request. In the context plot, medians and interquartile
ranges use the attempts still running at each request; they do not
represent a fixed cohort throughout.

Examples follow the selection rules fixed before collection: the lowest
seed with a local pass and external failure in each condition, the lowest
feedback seed with a diagnostic followed by a relevant edit and later pass,
and the lowest seed with an unsuccessful feedback repair. Each example
retains its final outcome. Figures~\ref{fig:teaser} and~\ref{fig:method}
remain illustrations, separate from these recorded attempts.

The public release contains the code and recorded measurements.
Raw trajectories and generated candidate sources are retained privately.

\section{Results}
\label{sec:results}

\begin{table*}[t]
  \caption{Migration outcomes and median editing costs. Full pass requires all 40 checks.}
  \label{tab:results}
  \centering
  \setlength{\tabcolsep}{4pt}
  \renewcommand{\arraystretch}{1.12}
  \begin{tabular*}{\textwidth}{@{\extracolsep{\fill}}lllrrrrr@{}}
    \toprule
    Model & Agent & Condition & Builds & Local pass & Full pass & Actions & Time (min) \\
    \midrule
    Qwen3-Coder-30B-A3B & mini-swe-agent & Baseline & 13/20 & 4/20 & 0/20 & 26.5 & 27.6 \\
     & & Feedback & 17/20 & 4/20 & 0/20 & 25.0 & 29.5 \\
    \addlinespace[4pt]
    Qwen3.8-27B (Q4), 128K & mini-swe-agent & Baseline & 18/20 & 18/20 & 18/20 & 54.0 & 25.7 \\
     & & Feedback & 20/20 & 18/20 & 18/20 & 45.5 & 26.1 \\
    \addlinespace[4pt]
    Qwen3.8-27B (Q4), 32K & mini-swe-agent & Baseline & 3/20 & 3/20 & 3/20 & 23.5 & 9.4 \\
     & & Feedback & 3/20 & 2/20 & 1/20 & 18.5 & 9.8 \\
    \addlinespace[4pt]
    Nemotron 3.5 Lightning & mini-swe-agent & Baseline & 2/20 & 2/20 & 0/20 & 56.0 & 4.5 \\
     & & Feedback & 4/20 & 1/20 & 0/20 & 56.0 & 6.3 \\
    \addlinespace[4pt]
    GPT-6 Astra & Codex & Baseline & 1/1 & 1/1 & 1/1 & 12 & 6.5 \\
     & & Feedback & 1/1 & 1/1 & 1/1 & 10 & 5.6 \\
    \addlinespace[4pt]
    Fable 5.1 & Claude Code & Baseline & 1/1 & 1/1 & 1/1 & 8 & 2.8 \\
     & & Feedback & 1/1 & 1/1 & 1/1 & 8 & 2.9 \\
    \bottomrule
  \end{tabular*}
  \par\smallskip
  {\footnotesize Qwen3.8 rows differ only in the served context window; all other
  settings are identical. Nemotron 3.5 Lightning is served at 128K. Native-agent rows
  are individual trials; action units and serving configurations differ.}
\end{table*}
\subsection{Final correctness}
Checker access does not increase the observed full-pass rate in any
comparison (Table~\ref{tab:results}). Qwen3-Coder passes 0/20 in each
condition, with a 95\% Wilson interval of 0--16.1\%. Qwen3.8 at 128K passes
18/20 in each, with an interval of 69.9--97.2\%. Nemotron 3.5 Lightning passes
0/20 in each. The observed difference is zero percentage points in those three
comparisons; at 32K Qwen3.8 passes 3/20 without and 1/20 with the checker.
Their intervals are 5.2--36.0\% and 0.9--23.6\%, respectively, with an
observed difference of $-10$ percentage points. These samples do not
establish equivalence or a reliable feedback advantage.

All 20 Qwen3-Coder and all 20 Nemotron seed pairs fail in both conditions. For Qwen3.8 at 128K,
16 pairs pass in both, two pass only in the baseline, and two pass
only with feedback. At 32K, three pass only in the baseline, one only
with feedback, and 16 fail in both. All 160 counted attempts receive final scores,
with no infrastructure exclusions, replacements, or final checker
execution errors. Source inventories, prompt hashes, model records,
and sampling settings match their respective frozen configurations.

For Qwen3-Coder, 13 baseline and 17 feedback patches compile, but this difference
does not carry through to complete correctness. The four local passes in each condition all
have at least one external-compatibility or required-rejection failure.
Thus, the common local round trip would accept eight final patches that
the contract-based evaluation rejects. This comparison uses the same
patch and fresh evaluation inputs for both kinds of check; it does not
rely on the agent's description of its tests.
The 32K Qwen3.8 batch adds one local-pass/external-fail patch and Nemotron
adds three, bringing the total to twelve across the local configurations.

\subsection{Qwen3-Coder failures}
Key handling blocks interoperability early. OpenSSL cannot load the
public key from any of the 30 built patches, while it loads the private
key from one feedback patch. Both directions involving independently
generated OpenSSL keys also fail for every built patch
(Table~\ref{tab:properties}). The independent OpenSSL key-generation and
signing setup checks pass in all 30 built evaluations.

\begin{table}[t]
  \caption{Qwen3-Coder checks: pass/fail/unassessed.}
  \label{tab:properties}
  \centering\small
  \setlength{\tabcolsep}{4pt}
  \begin{tabularx}{\columnwidth}{@{}Xcc@{}}
    \toprule
    Check & Baseline & Feedback \\
    \midrule
    Application generates keys & 6/7/7 & 7/10/3 \\
    Application signs a message & 5/8/7 & 4/13/3 \\
    Application verifies its signature & 4/9/7 & 4/13/3 \\
    OpenSSL loads application public key & 0/13/7 & 0/17/3 \\
    OpenSSL loads application private key & 0/13/7 & 1/16/3 \\
    Application verifies a foreign signature & 0/13/7 & 0/17/3 \\
    Application signs with an OpenSSL key & 0/13/7 & 0/17/3 \\
    \bottomrule
  \end{tabularx}
\end{table}

A failed prerequisite can produce several later failures. If OpenSSL
cannot read an application key, it cannot create the wrong-context
signature needed for a rejection test. A later failure to read that
missing signature does not show that the application accepts an
incorrect context. We retain the raw verdicts for the primary outcome
and trace their prerequisites before attributing semantic errors.

Build failures also have different causes. Four baseline patches and
one feedback patch stop at compiler errors such as undefined APIs or
unused variables. Two patches in each condition import CIRCL components
whose transitive dependencies are absent from the offline proxy.
A further baseline patch deletes \texttt{go.sum} and adds a replacement
path into its temporary module cache; its final read-only build stops
when Go tries to update checksums. These are failures under the recorded
build environment, not evidence about signature correctness.

The ten failed builds produce only the four preflight rows. Their missing
candidate binary is recorded as a dependency failure, with 36 subsequent
checks unassessed. Built patches that do not produce a signature can
also omit the signature-size row. No omitted row is treated as a pass,
and every built patch has observed failures beyond any omitted row.

\subsection{Qwen3.8 at 128K}
All 18 baseline patches that compile pass every final check. All 20
feedback patches compile, of which 18 pass. No final patch in this batch
passes the local round trip while failing an external requirement.

The four unsuccessful attempts stop with incomplete migrations. Baseline
seeds 7 and 12 leave no migration source to build; feedback seeds 4 and
19 compile but record 29 and 24 passed checks, respectively. The first
three stop after consecutive output-limit responses that the action
parser rejects. Feedback seed 19 reaches the step limit. These are
observed stopping mechanisms, not evidence that longer runs would
necessarily succeed. All four remain in the denominator.

There are no request errors or context-limit stops in the counted
Qwen3.8 batch. Output-limit stops occur in otherwise successful attempts
as well: baseline seed 18 ends on a format error, but its saved patch
passes all 40 checks. The final evaluator therefore assesses the patch
left by the agent, independently of its stopping message.

Eighteen feedback attempts invoke the checker, twelve twice and six
once. Every first call already passes all 40 checks, as do the later
calls and final evaluations. The two unsuccessful feedback attempts
never call it. This association does not show that checking caused
success: the checked patches were already correct on the first call.
No checker-diagnosed repair or local-pass/external-fail example is
observed in this batch under the selection rules.

\subsection{Context window}
\label{sec:context}
Reducing only Qwen3.8's served window from 128K to 32K lowers full passes
from 36/40 to 4/40 across the two conditions. Matching condition and seed,
33 attempts pass at 128K and fail at 32K, three fail at both, three pass
at both, and one passes only at 32K. Six patches compile at 32K,
compared with 38 at 128K.

\begin{figure}[t]
  \centering
  \includegraphics[width=\columnwidth]{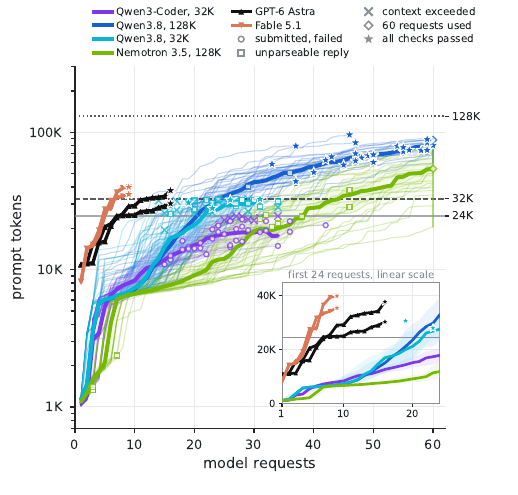}
  \caption{Prompt growth across model requests. Faint traces show attempts;
  solid lines and bands show medians and interquartile ranges among
  continuing local attempts. Stars mark full passes. At request 60,
  endpoints are grouped by model and outcome; markers and vertical bars
  show medians and ranges. The 24K line reserves 8K for a response in a 32K window.}
  \label{fig:context}
\end{figure}

Retained history competes with the next response for the context window
(Figure~\ref{fig:context}). As the prompt grows, the server can shorten
the response allowance; an oversized request is eventually rejected.
At 32K, 21 attempts stop on context errors and 19 on repeated unparseable
responses. Nine of the latter end with three consecutive responses that
reach their output limit with reasoning but no command. This is one
observed route to a format stop, not an explanation for every such stop.
None submits voluntarily, compared with 27/40 at 128K. Nevertheless,
four saved patches pass all checks after 18--26 executed actions.

The ablation shows that the served window substantially affects this
Qwen3.8 configuration. It does not establish a minimum context needed
to solve the task. Nor does it isolate Qwen3-Coder's failure: 32 of its
40 attempts submit voluntarily, and a larger-window version of that
configuration was not tested.

\subsection{Nemotron outcomes}
Nemotron 3.5 Lightning completes 0/20 migrations per condition at 128K.
Twenty-eight attempts exhaust the 60-request budget and twelve stop on
repeated unparseable responses. There are no context-limit errors; the
largest recorded prompt contains 78{,}455 tokens, below the 131{,}072-token
window. Six patches compile and pass 12--25 checks. Thirty attempts
execute commands but leave a failed build, including imports unavailable
in the offline module cache. Four execute no command before a format
stop. All forty remain in the denominator.

The three local passes all fail external requirements. Only two feedback
attempts invoke the checker, once each, and both checks fail. The median
attempt executes 56 actions and takes 5.3 minutes across conditions.
These runs reach request or format limits well before the one-hour cap;
their speed does not imply progress toward a correct migration.

\subsection{Feedback and cost}
For Qwen3-Coder, sixteen of the 20 feedback attempts invoke the checker: ten once and
six twice. Four never use it. Checker access therefore measures the
effect of making diagnostics available, not an intervention that forces
an agent to use or act on them. The baseline remains free to run
OpenSSL directly.

\begin{table}[t]
  \caption{Local-agent costs: median (IQR).}
  \label{tab:cost}
  \centering\small
  \setlength{\tabcolsep}{4pt}
  \begin{tabularx}{\columnwidth}{@{}Xcc@{}}
    \toprule
    Measure & Baseline & Feedback \\
    \midrule
    \multicolumn{3}{l}{\textit{Qwen3-Coder}} \\
    Executed actions & 26.5 (8.25) & 25.0 (7.50) \\
    Development checker calls & 0 (0) & 1 (1) \\
    Attempt time (min) & 27.6 (17.1) & 29.5 (15.6) \\
    Final scoring time (s) & 11.04 (11.19) & 11.29 (0.58) \\
    \addlinespace
    \multicolumn{3}{l}{\textit{Qwen3.8, 128K}} \\
    Executed actions & 54.0 (16.5) & 45.5 (12.5) \\
    Development checker calls & 0 (0) & 2 (1) \\
    Attempt time (min) & 25.7 (6.4) & 26.1 (5.0) \\
    Final scoring time (s) & 10.93 (0.37) & 10.87 (0.34) \\
    \addlinespace
    \multicolumn{3}{l}{\textit{Qwen3.8, 32K}} \\
    Executed actions & 23.5 (3.25) & 18.5 (6.00) \\
    Development checker calls & 0 (0) & 0 (0) \\
    Attempt time (min) & 9.4 (4.1) & 9.8 (5.9) \\
    Final scoring time (s) & 0.11 (0.01) & 0.11 (0.01) \\
    \addlinespace
    \multicolumn{3}{l}{\textit{Nemotron 3.5 Lightning}} \\
    Executed actions & 56.0 (27.0) & 56.0 (9.25) \\
    Development checker calls & 0 (0) & 0 (0) \\
    Attempt time (min) & 4.5 (5.0) & 6.3 (4.0) \\
    Final scoring time (s) & 0.11 (0.01) & 0.11 (0.01) \\
    \bottomrule
  \end{tabularx}
\end{table}

For Qwen3-Coder, median attempt time is 27.6 minutes for the baseline and 29.5 minutes
with feedback (Table~\ref{tab:cost}). This difference includes model
generation and all development actions; it is not an estimate of
checker overhead alone. Final scoring is outside the editing budget; its duration includes
the early exits of failed builds.

Five Qwen3-Coder baseline attempts and three feedback attempts stop when the model
request exceeds the configured context window. They remain scored using
the final workspace. The remaining attempts submit voluntarily.
The one-hour limit therefore does not explain every failure. Submission also does not show that known errors were repaired.

At 128K, Qwen3.8 uses fewer median executed actions with feedback (45.5 versus
54.0), but median elapsed time is similar (26.1 versus 25.7 minutes).
Action counts include checker calls and omit rejected responses;
elapsed time also includes reasoning and model generation. These
descriptive differences do not establish faster convergence, and the
different serving configurations preclude a direct model-speed comparison.
At 32K, no Qwen3.8 attempt calls the checker before stopping. A zero
median also hides Nemotron's two calls; the traces are needed to
distinguish unused feedback from unsuccessful repairs.

\subsection{Native outcomes}
All four scored native trials compile and pass all 40 final checks
(Table~\ref{tab:results}). Both baselines succeed without checker
feedback. The task is therefore solvable within the supplied
environment, but these cases do not isolate why the Qwen3-Coder configuration
failed: the model, interface, and context handling also change.

Codex uses two development checks. The first records 20 passes and
19 failures; the second and final evaluations pass all 40 checks.
Claude's single development check already passes all 40. Thus the
feedback traces differ, and neither pair establishes a feedback advantage
or a reliable success rate from one trial per condition.

One additional Claude baseline attempt loses its connection after six
actions and is archived before scoring. Inspection finds an unintended
idle timeout in the proxy, which is corrected before a fresh replacement;
no tunnel log proves it caused that interruption. Its exclusion is an
operator-classified exception to the verified-outage rule. Separately, Codex's
model-accounting rule initially mistakes its auxiliary approval model
for the coding model. Correcting that rule retains the original candidate,
whose post-run and scored source hashes match. These amendments and the
interrupted record are retained; the task and final evaluator are unchanged.

\section{Repair Analysis}
\label{sec:analysis}
\subsection{Local agreement, incompatible keys}
In Qwen3-Coder, baseline seed 5 and feedback seed 4 are the first eligible local-pass,
external-fail examples. Both write CIRCL's packed key bytes directly
inside PEM blocks labeled \texttt{PRIVATE KEY} and \texttt{PUBLIC KEY}.
Their loaders reverse that operation with \texttt{UnmarshalBinary}.
These two sides agree, so local signing and verification pass. The
stored bytes do not contain the required PKCS\#8 and SubjectPublicKeyInfo
structures, and OpenSSL rejects both key files.

\begin{figure}[t]
  \centering
  \setlength{\fboxsep}{8pt}
  \fcolorbox{navy!25}{pale}{%
  \begin{minipage}{\dimexpr\columnwidth-2\fboxsep-2\fboxrule\relax}
    \small\raggedright
    \textcolor{navy}{\textbf{Baseline 5 and Feedback 4}}\par
    Packed key bytes $\rightarrow$ PEM wrapper $\rightarrow$
    packed-key loader.\par\smallskip
    \textbf{Local check:} sign and verify pass.\par
    \textbf{External check:} OpenSSL rejects the keys.\par
    \textbf{Final outcome:} 22 pass, 18 fail in each attempt.\par\medskip
    \textcolor{navy}{\textbf{Feedback 7: a partial repair}}\par
    \textbf{Diagnostic:} local signature verification fails.\par
    \textbf{Edit:} use \texttt{SignTo} with the required context.\par
    \textbf{Recheck:} local verification passes.\par
    \textbf{Final outcome:} 22 pass, 18 fail; key formats
    and other requirements remain unresolved.
  \end{minipage}}
  \caption{Recorded Qwen3-Coder errors and a partial repair.}
  \label{fig:qualitative}
\end{figure}

Each final evaluation records 22 passes and 18 failures
(Figure~\ref{fig:qualitative}). Those failures include malformed-length
signatures returning the invalid-signature code and deterministic signing
where the contract requires randomized signing. The wrong-context and
empty-context rejection rows, however, depend on signatures that OpenSSL
cannot construct with the incompatible keys. They do not establish
independent context-rejection defects.

Feedback seed 4 receives the key-loading diagnostics, reads library
documentation, and then tests local key generation, signing, and
verification. Those local tests succeed. The final migration source is
unchanged from the checked version, leaving the external failures
unresolved. The example shows why repeating a successful local test
cannot settle a key-format mismatch identified by another implementation.

\subsection{A repaired context mismatch}
Feedback seed 7 is the lowest seed with a diagnostic, relevant edit,
and later pass on the diagnosed check. Its first checked version calls
\texttt{sk.Sign} without a context, while verification supplies
\texttt{filesigner/v1}. In the pinned CIRCL version, that signing method
uses an empty context~\cite{circlmldsa}. The checker reports failed local
verification. The agent reproduces the failure and replaces the signing
call with \texttt{SignTo}, supplying the required context explicitly.

The second development check passes local verification and three related
verification checks. Between the two checked source versions, the signing
call and signature-buffer allocation are the only functional changes.
The source remains unchanged at final evaluation, which again records
22 passes and 18 failures. OpenSSL key loading and the randomized-signing
requirement still fail. This repair is observed within one feedback
attempt and does not establish a causal advantage over the baseline.

This observed context mismatch differs from the teaser's illustrative
shared mistake. Here, signing and verification use different contexts,
so even a local test can expose it. The recorded examples in which local
agreement hides an external failure concern key encodings.

\subsection{An unsuccessful repair}
Feedback seed 1 is the first unsuccessful feedback repair. It binds the
return values of key generation in reverse order, then passes the
resulting public-key object to a private-key marshaler. The checker
reports the key-generation failure. Two later source rewrites retain
the faulty binding and marshaling call. A combined build-and-check shell
command also fails to invoke the host's standalone checker action.

The final source compiles but cannot generate its keys. Evaluation records
9 passes, 30 failures, and an unassessed signature-size row. Many failures
follow from missing keys or signatures; they are not 30 independently
identified bugs. The trace shows that receiving diagnostics and making
subsequent edits is insufficient evidence that the reported defect was
repaired.

\section{Discussion}
Independent checks expose errors that local round trips miss, but
diagnostics alone need not supply the knowledge or editing strategy
needed for repair.

The key examples suggest verifying key exchange before signature and
rejection behavior. Otherwise, one serialization error can produce
several dependent failures. Diagnostics that identify these dependencies
may help agents prioritize repairs; this study does not test that
interface.

Checker availability leads to different behaviors. Qwen3-Coder receives
failing checks but often leaves the underlying defect unresolved.
Qwen3.8 at 128K checks patches that already pass; at 32K it never calls
the checker. Nemotron uses it in only two attempts. None of the four
comparisons shows a higher completion rate with feedback. This result
concerns optional access, not the effect of requiring a check or a repair.

The context ablation also changes the interpretation of failure. With
the same model, weights, harness, and budgets, reducing Qwen3.8's window
lowers completion from 36/40 to 4/40. Evaluating that model only at 32K
would miss most of its successful migrations. Yet more context is not
a general remedy: Nemotron fails without reaching its window, and
Qwen3-Coder usually submits before a limit intervenes. The native
baselines establish feasibility under their own configurations, without
isolating which model or interface feature makes them succeed.

\section{Limitations}
The study covers one Go file signer and one ML-DSA parameter set.
Repeated attempts do not establish performance across repositories or
production migrations; the four native trials cannot estimate success
rates. The Qwen3.8 configuration was selected after inspecting excluded
setup pilots. Across model configurations, only the Qwen3.8
window ablation changes a single serving parameter. Other configurations
differ in quantization, sampling, architecture, or inference engine, so cross-model differences cannot be
attributed to the model alone. OpenSSL is the only external
implementation, and the task's key-format requirements are a substantial
part of its difficulty.

The offline proxy excludes dependencies outside the supplied cache,
including those required by some CIRCL import paths. One model repeatedly
imports an unavailable package, so this constraint contributes to its
failures. Near the context limit, the runtime can reduce the response
allowance before rejecting an oversized request; reasoning retention and
response parsing therefore affect which limit ends a run. Final builds
use read-only snapshots, retaining workspaces after context, output, or
step limits. These shared constraints affect observed success and limit
comparisons with unrestricted coding environments.

The checker exercises selected contract behaviors. Known-answer tests
are not integrated; neither a local pass nor a full checker pass would
establish FIPS conformance, side-channel resistance, or deployment safety.

\section{Conclusion}
We evaluate post-quantum migration against a contract and
independent key and signature exchange. Across 160 attempts in four
local-agent configurations, optional checker access does not increase
observed completion. Twelve patches pass local verification but fail external
requirements. Reducing only Qwen3.8's context window lowers full passes
from 36/40 to 4/40; Qwen3-Coder and Nemotron complete none. All four
native-agent trials pass, including both baselines. Migration evaluation
must distinguish local agreement from interoperability and report harness
and serving limits. Diagnostics can expose defects without ensuring repair.

\bibliographystyle{IEEEtran}
\IEEEtriggeratref{12}
\bibliography{refs}
\clearpage
\appendices
\section{Prompts}
\label{app:prompts}
The local comparisons use the task prompt below. The feedback condition
inserts the additional bullet before the build/test/repair instruction;
the baseline omits it. Line wrapping is adjusted for the column. The
agent also receives the contract and the harness's command-format
instructions. This box reproduces the task prompt, not the full context.

\begingroup
\setlength{\fboxsep}{6pt}
\noindent\fcolorbox{blue!25}{pale}{\begin{minipage}{\dimexpr\columnwidth-2\fboxsep-2\fboxrule\relax}
\footnotesize\raggedright
You are working in /work, a Go module named \texttt{filesigner}.

\smallskip
\path{cmd/rsa-signer/main.go} is a command-line file signer that uses
RSA-2048 with PKCS\#1 v1.5. Migrate it to ML-DSA-44 by creating
\path{cmd/mldsa-signer/main.go}: a program with the same three commands
that follows CONTRACT.md exactly. CONTRACT.md is the specification;
where it and the RSA code disagree, the contract wins. Do not modify
\path{cmd/rsa-signer}.

\smallskip
Your environment:
\begin{list}{--}{\leftmargin=1em\itemsep=2pt\parsep=0pt\topsep=3pt}
\item There is no network access.
\item Go 1.27.1 is on PATH. \path{github.com/cloudflare/circl} v1.6.5,
already required in go.mod, is available from an offline module proxy.
No other third-party modules are available.
\item Build with: \texttt{go build -o /tmp/filesigner-mldsa}
\path{./cmd/mldsa-signer}
\item OpenSSL 3.5.8, with ML-DSA-44 support, is on PATH as
\texttt{openssl} while you develop. It is not available to your program
during final scoring: the submitted program runs without OpenSSL, so it
must not call it.
\end{list}
Before you submit, build your program, test it against the requirements
in CONTRACT.md, and repair every failure you find. Submit only after that.

\smallskip
You may run at most 60 commands. After you finish,
\path{cmd/mldsa-signer} will be built from /work and tested independently
against CONTRACT.md, using inputs you have not seen.
\end{minipage}}
\endgroup

\smallskip
\noindent\textbf{Additional feedback bullet (verbatim).}
\begin{quote}\footnotesize
A checker is available. Send the command \texttt{check}, exactly that and
nothing else, to build the current /work and run the contract checker
against it. It prints every check's name, its result (PASS, FAIL, DEPMISS
or ERROR) and a short detail. Each run uses new inputs. Each \texttt{check}
counts as one of your commands.
\end{quote}
The prompt says 60 commands; the harness counts model requests, including
rejected replies. The frozen wording is retained here.

\smallskip
\noindent\textbf{Native-agent changes.}
The native prompt describes blocked downloads rather than an absent
network, counts shell commands and file edits toward 60 tool actions,
and treats ending the turn as submission. Its checker bullet asks for
the shell command \texttt{check}. Both native conditions receive the
same build/test/repair instruction and contract.

\smallskip
\noindent\textbf{Native settings.}
Codex CLI 0.154.0 uses \texttt{gpt-6-astra}; Claude Code 2.1.272 uses
\texttt{claude-fable-5-1}. Both use high reasoning effort, one hour and
60 tool actions. Their context handling follows the CLI, and sampling
seeds are not exposed. Claude's command timeout is 120 seconds;
Codex chooses its command timeouts. These cases are not repetitions of
the local-agent setup.

\newpage
\section{Settings and execution}
\label{app:settings}
Table~\ref{tab:appendix-settings} collects the local serving settings.
All four configurations use mini-swe-agent 2.4.6 and LiteLLM 1.100.1,
with paired seeds 1--20 per condition. The model and artifact references
identify the pinned weight revisions~\cite{qwen3coder,qwen38,qwen38gguf,nemotron,nemotrongguf}.

\par\smallskip\noindent
\begin{minipage}{\columnwidth}
\makeatletter\def\@captype{table}\makeatother
\centering
\caption{Local serving settings. Qwen3.8 has two context variants.}
\label{tab:appendix-settings}
\footnotesize
\setlength{\tabcolsep}{3pt}
\begin{tabularx}{\columnwidth}{@{}lXXX@{}}
\toprule
Setting & Qwen3-Coder & Qwen3.8 & Nemotron \\
\midrule
Weights & FP16 & Q4\_K\_M & Q4\_0 \\
Context & 32K & 128K / 32K & 128K \\
MTP head & None & Q8\_0 & Q4\_0 \\
Engine & vLLM & llama.cpp & llama.cpp \\
Revision & 0.24.0 & \texttt{4a89937} & \texttt{4a89937} \\
Temperature & 0.7 & 1.0 & 1.0 \\
Top-$p$ & 0.8 & 0.95 & 0.95 \\
Top-$k$ & 20 & 20 & 0 (off) \\
Min-$p$ & Not set & 0 & 0 \\
Repeat penalty & 1.05 & Not set & Not set \\
Parallel attempts & 1 & 2 & 2 \\
\bottomrule
\end{tabularx}
\end{minipage}
\par\medskip

\noindent\textbf{Shared local limits.}
Each attempt allows 60 model requests and 3,600 seconds of editing.
Each response allows up to 8,192 output tokens; reasoning uses the same
allowance. Commands have a 120-second limit. The Qwen3.8 and Nemotron
clients allow 1,500 seconds per model request. Their thinking mode and
retention of prior reasoning are enabled. Three consecutive unparseable
responses stop a local attempt. Final scoring lies outside the editing
budget; its build has a 300-second limit and each candidate invocation
has a 30-second limit.

Figure~\ref{fig:verdict} classifies exit-code checks. Length, DER equality
and signature inequality are checked directly. Missing prerequisites stop
cryptographic testing; failed builds remain unsuccessful attempts.

\par\medskip\noindent
\begin{minipage}{\columnwidth}
\makeatletter\def\@captype{figure}\makeatother
\centering
\includegraphics[width=\columnwidth]{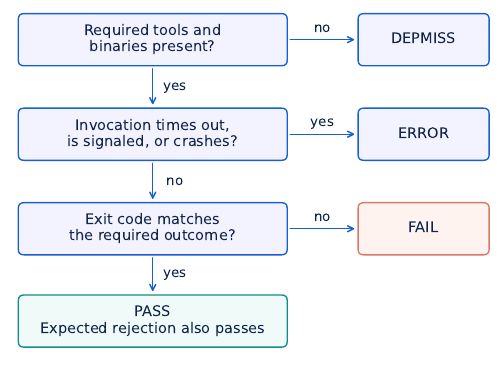}
\caption{Exit-code checks distinguish expected rejection from crashes and missing prerequisites.}
\label{fig:verdict}
\end{minipage}
\par

\clearpage
\makeatletter
\twocolumn[{
\begin{minipage}{\textwidth}
\def\lst@makecaption#1#2{\@makecaption{#1}{#2}\vspace{8pt}}
\begin{lstlisting}[language=Python,
  caption={Local attempt and final scoring (pseudocode)},
  label={alg:attempt},captionpos=t,
  basicstyle=\fontsize{9}{10.5}\selectfont\ttfamily,
  keywordstyle=\color{blue},stringstyle=\color{teal},
  numbers=left,numberstyle=\scriptsize\color{navy!60},numbersep=10pt,
  xleftmargin=22pt,frame=tb,rulecolor=\color{blue!35},
  framesep=6pt,showstringspaces=false,keepspaces=true,
  columns=fullflexible,aboveskip=0pt,belowskip=6pt]
workspace, history = fresh_start(condition)
requests, format_errors = 0, 0
while requests < 60 and time_left():
    requests += 1
    reply = query_model(history)
    if reply.stopped:
        break
    command = parse_command(reply)
    if command is None:
        history.extend(parser_feedback(reply))
        format_errors += 1
        if format_errors == 3:
            break
        continue
    format_errors = 0
    if condition == "feedback" and command == "check":
        result = check_patch(workspace, fresh_inputs=True)
    else:
        result = execute(command, workspace)
    history.extend([reply, result])
    if result.submitted:
        break

patch = snapshot(workspace)
result = final_score(patch, read_only=True, fresh_inputs=True)
full_pass = result.build_ok and result.passed_checks == 40
\end{lstlisting}
\small
\texttt{final\_score} rebuilds a read-only snapshot and checks it through isolated calls.
Stopped attempts retain their final workspace; infrastructure replacements follow the main-text rules.
\end{minipage}
\vspace{8pt}
}]
\makeatother
\section{Checker coverage}
\label{app:coverage}
Table~\ref{tab:coverage} groups all 40 rows in execution order, including
prerequisite and helper operations. They are not 40 independent security
properties: a key-loading failure can cause several later failures.

\par\smallskip\noindent
\begin{minipage}{\columnwidth}
\makeatletter\def\@captype{table}\makeatother
\centering
\caption{Coverage of the 40 checker rows.}
\label{tab:coverage}
\footnotesize
\begin{tabularx}{\columnwidth}{@{}rX@{}}
\toprule
Rows & Required behavior \\
\midrule
1--4 & OpenSSL supports ML-DSA-44; candidate and RSA binaries exist. \\
5--8 & Generate keys, sign, verify locally; signature length is 2,420 bytes. \\
9--19 & Load keys in OpenSSL and identify ML-DSA-44; cross-sign and verify using keys generated by each implementation. \\
20--27 & Reject changed message, wrong key, wrong/empty context and flipped bit; includes three helper calls. \\
28--32 & Treat short/long signatures, garbage public key and RSA key as malformed; includes RSA key generation. \\
33--35 & Reload keys; sign and verify again locally and with OpenSSL. \\
36--37 & Match derived/public SPKI DER; accept CRLF public-key PEM. \\
38--40 & Create a second signature; require correct sizes, distinct bytes and successful verification. \\
\bottomrule
\end{tabularx}
\end{minipage}
\par\medskip

Each score records a fresh seed and draws a message of 1--4,096 bytes.
Message tampering flips one bit. Signature tampering flips a bit in the
first 2,336 bytes, preserving hint encoding to test cryptographic rejection
rather than malformed-input rejection.

Positive controls test false alarms: corruption must change an already-set
byte, ordinary rejection text must not count as a crash, and the unchanged
reference must pass.

\newpage
\section{Validation controls}
\label{app:controls}
All 16 controls in Table~\ref{tab:controls} meet their expected outcome
on the v4 checker and through isolated scoring. Three must remain clean;
thirteen expose a defect or missing dependency. They are excluded from
agent results.

\par\smallskip\noindent
\begin{minipage}{\columnwidth}
\makeatletter\def\@captype{table}\makeatother
\centering
\caption{Validation controls and expected checker behavior.}
\label{tab:controls}
\footnotesize
\begin{tabularx}{\columnwidth}{@{}Xl@{}}
\toprule
Control & Expected outcome \\
\midrule
Empty signing context & FAIL \\
Prehash instead of pure ML-DSA & FAIL \\
Verification accepts everything & FAIL \\
Malformed input returns rejection code & FAIL \\
Raw public key instead of SPKI & FAIL \\
Verification panics & ERROR \\
Verification hangs & ERROR \\
Deterministic signing & FAIL \\
Valid signature with byte 100 = 0x01 & Clean \\
Key generation hangs & ERROR \\
Signing succeeds but writes nothing & Randomization FAIL \\
Valid rejection mentions runtime error & Clean \\
Failure detail contains special characters & FAIL, valid JSON \\
Embedded and published keys disagree & FAIL \\
OpenSSL missing & DEPMISS \\
Unmodified reference & Clean \\
\bottomrule
\end{tabularx}
\end{minipage}
\par\medskip

Clean controls have no FAIL, ERROR or DEPMISS rows. Faults must produce
the expected category; a timeout cannot substitute for a semantic failure.
All outputs must be valid JSON, including details with quotes, backslashes
and tabs.

\end{document}